\documentclass[]{article}

\usepackage{amsfonts}
\usepackage{dsfont}
\usepackage{amsmath}
\usepackage{verbatim}
\usepackage{amsthm}
\usepackage{enumerate}
\usepackage{bm}

\theoremstyle{plain}
\newtheorem{definition}{Definition}

\newtheorem{lemma}{Lemma}

\theoremstyle{definition}
\newtheorem{remark}{Remark}

\newtheorem{problem}{Problem}

\usepackage{algorithmic}
\usepackage{algorithm}

\usepackage{fancyvrb}
\usepackage[margin=20mm, nohead]{geometry}
\DefineVerbatimEnvironment{code}{Verbatim}{fontsize=\small}
\DefineVerbatimEnvironment{example}{Verbatim}{fontsize=\small}

\usepackage{listings}
\title{A macro placer algorithm for chip design}

\author{Endre Cs\'oka, Attila De\'ak}

\begin{document}

\maketitle

\begin{abstract}
There is a set of rectangular macros with given dimensions, and there are wires connecting some pairs (or sets) of them. We have a placement area where these macros should be placed without overlaps in order to minimize the total length of wires. We present a heuristic algorithm which utilizes a special data structure for representing two dimensional stepfunctions. This results in fast integral computation and function modification over rectangles. Our heuristics, especially our data structure for two-dimensional functions, may be useful in other applications, as well.
\end{abstract}

\section{Introduction}
A chip is composed of basic elements called cells, circuits, boxes or modules.
They usually have a rectangular shape, contain several transistors and internal connections, and have (at least two) fixed pins.
There is a netlist describing which pin should be connected to which other pins.
The goal is to place the cells legally -- without overlaps -- in the chip area so as to minimize the total (weighted) length of the wires connecting the pins.
This problem is also called the VLSI placement problem.

Finding the optimum is NP-hard, therefore, we present a heuristic algorithm based on primal-dual optimization inspired by the Hungarian Algorithm \cite{magyar} for the minimum weight maximum matching problem.
Namely, we use a cost function as dual function on the placement area, and we are looking for a placement minimizing the sum of the total netlength and the total costs of the areas covered by the macros. We try to find a non-negative cost function by which an almost optimal placement is legal even if we allow overlaps, and costs are counted with multiplicity.
We will use an iterated algorithm on the space of primal-dual pairs based on the following two steps:

\begin{enumerate}[1.]
\item
For every overlap, we increase the cost function under intersecting areas.

\item
We try to find a better placement with respect to the new cost function.

\end{enumerate}
%

We tried to focus on typical instances in practice, and we found that these have the following properties.
\begin{enumerate}
\item The placement area is not very large compared to the total size of the macros, but it is still easy to find a legal placement.
\item There are about a few hundreds of macros and every macro is contained in at most $10$ nets.
\item Most of the nets connect two, sometimes three, and rarely more than three macros to each other.
\end{enumerate}
Our method is optimized for such inputs.

\bigskip

This paper is organized as follows. In Section \ref{sec.macro.placement}, we introduce some notations and give a formal definition of the macro placement problem. In Section \ref{sec.basic.tools}, we describe the basic idea behind our algorithm and in Section \ref{sec.global.placer}, we present the algorithm. In Section \ref{sec.heuristics}, we describe some additional heuristics used in our placer.
%
%
%
%
\section{The macro placement problem}
\label{sec.macro.placement}
Now we give a formal definition of the simplified macro placement problem.
Let us denote by $\cal M$ the set of macros.
We assume that all pins of each macro are in the center of the macro.
The place of a macro is identified with the place of its center pin.
For a macro $M$, denote its horizontal and vertical size by $size_x(M), size_y(M)$, respectively.
For a macro $M$ at $(x,y)$, we denote the area occupied by $M$ by
\begin{equation*}
S(M,(x,y))=
\left(x-{size_x(M)\over 2},\ x+{size_x(M)\over 2}\right]\times
\left(y-{size_y(M)\over 2},\ y+{size_y(M)\over 2}\right].
\end{equation*}
A net $N$ is a subset of the macros that are connected.
\begin{definition}
A netlist is a pair $({\cal M,N})$ where
$\cal M$ is a finite set of macros and $\cal N\subseteq {\cal P}({\cal M})$ is a set of subsets of ${\cal M}$.
\end{definition}
One can think of $\cal N$ as a hypergraph on $\cal M$, where each $N\in {\cal N}$ is a hyperedge.
We assume $|N|\geq 2$ for each $N\in {\cal N}$.
\begin{definition}
The placement area is a rectangle denoted by $\cal A$. This contains a set of rectangular blockages ${\cal B}$ (where $\forall B\in {\cal B},\, B\subset {\cal A}$). The sides of all rectangles are parallel to the axis.
\end{definition}
\noindent
A blockage is a part of the placement area where no macro can be placed.
\begin{definition}
A placement is a map $p:{\cal M}\mapsto {\cal A}$.
The placement $p$ is legal if all of the followings hold.
\begin{itemize}

\item
Every macro $M\in {\cal M}$ is placed in the placement area:
\begin{equation*}
S\big(M,p(M)\big)\subseteq {\cal A}.
\end{equation*}

\item
The places of any two macros $M, M' \in {\cal M}$ are disjoint:
\begin{equation*}
S\big(M,p(M)\big)\cap S\big(M',p(M')\big)=\emptyset.
\end{equation*}
 
\item
None of the macros $M\in {\cal M}$ are placed on a blockage $B\in {\cal B}$:
\begin{equation*}
S\big(M,p(M)\big)\cap B=\emptyset.
\end{equation*}

\end{itemize}
\end{definition}
The macros have to be placed in the given orientation, these cannot be rotated.
Let $({\cal M, N})$ be a netlist and $p$ a legal placement to the placement area $\cal A$ with blockages $\cal B$.
Define $p$ on the set of nets $\cal N$ as follows.
For $N=\{M_1,M_2,\dots , M_k\}\in {\cal N}$, let $p(N)=\big(p(M_1),p(M_2),\dots ,p(M_k)\big)$.
We have a function ${\cal L}:{\cal A}^2\cup {\cal A}^3\cup {\cal A}^4\cup \dots \mapsto \mathds{R}^+$ which evaluates the length of a net.
${\cal L}$ is also called the net (or netlength) model.
One commonly used net model is the bounding-box model:
\begin{equation} \label{eqn.bounding.box}
BB \Big((x_1,y_1),(x_2,y_2),\dots ,(x_k,y_k)\Big)=
\max_{i}\{x_i\}-\min_{i}\{x_i\}+
\max_{i}\{y_i\}-\min_{i}\{y_i\}
\end{equation}

This is the half perimeter of the smallest rectangle with sides parallel to the axis, containing all pins of the macros contained in the net $N$.

\medskip
\noindent
{\bf The Simplified Placement Problem}:\\
\noindent
Given a netlist $({\cal M},{\cal N})$, a placement area $\cal A$, the set of blockages $\cal B$ and a net model $\cal L$,
find a legal placement $p:{\cal M}\mapsto {\cal A}$ which minimizes the total netlength:
$$\sum_{N\in {\cal N}}{\cal L}\big(p(N)\big).$$


\section{Basic tools of the placer}
\label{sec.basic.tools}
The initial problem is to place macros in the placement area (avoiding the blocked areas) so that the total netlength is minimal (or close to the minimum).
As finding the optimum is NP hard, we present a heuristic algorithm with ${\cal O}\big(\log(n)\log(m)s\big)$ running time, where the placement area is a discrete $n\times m$ grid and we run the algorithm for $s$ rounds.

We introduce our algorithm in several steps.

\begin{problem} We have a set of macros $\cal M$ and disjoint slots $\cal A$.
There is a cost function $c: {\cal M}\times {\cal A}\mapsto \mathds{R}^+$ which assigns costs to every possible macro-slot assignment.
Find an injective assignment $p:{\cal M}\mapsto {\cal A}$ with minimum total cost
\begin{equation*}
\sum_{M\in {\cal M}}c\big(M,\ p(M)\big).
\end{equation*}
\end{problem}

\noindent
{\bf Solution.}
This scenario can be represented by a bipartite weighted graph.
The two set of points are $\cal M$ and $\cal A$ and the cost of an edge $(M,A)$ is $c(M,A)$.
The task is to cover $\cal M$ with a minimum cost maximum matching.
This is a well known optimisation problem, and it is usually solved by primal-dual methods (e.g. the Hungarian Method \cite{magyar}).
However, we choose the following method instead, because we will generalize this in the later steps.
We try to find a primal-dual solution by the following market simulation: if there is an area which is the best possible choice for at least two macros, then we increase the cost of that area.

\bigskip
\begin{problem}
For a macro $M\in {\cal M}$ and a given netlist $\cal N$, let $E(M)$ be the set of nets containing $M$:
\begin{equation*}
E(M)=\big\{N\in {\cal N}\ \big|\ M\in  N\big\},
\end{equation*}
and let $N(M)$ be the set of its neighbors:
\begin{equation*}
N(M)=\big\{M'\in {\cal M}\ \big|\ \exists N\in E(M):\ M'\in N\big\}.
\end{equation*}
Let ${\cal L}(N)$ be the netlength model.
Given a set of macros ${\cal M}$, disjoint slots $\cal A$ and netlist $\cal N$,
find an injective assignment $p:{\cal M}\mapsto {\cal A}$ which minimizes
\begin{equation*}
\sum_{N\in {\cal N}}{\cal L}\big(p(N)\big).
\end{equation*}

\end{problem}

\noindent
{\bf Solution.}
We try to use the solution of Problem 1, where the cost of the placement of one macro is replaced by its marginal contribution to the total cost.
Now, the cost of one macro depends on the placement of its neighbors, but there are not too many neighbors, therefore, we expect the cost function to change rather slowly.
This allows us to use a variation of the above described market simulation:

\begin{enumerate}
\item{
Take an arbitrary macro-area assignment $p$.}

\item{\label{problem2.2}
For each macro $M\in {\cal M}$, fix the other macros at their current position.
For every $A\in {\cal A}$, we get a placement $p_{(M,A)}$ from $p$ by changing the assignment of the macro $M$ to $A$.
Let us define the marginal contribution of $M$ placed at $A$ by
\begin{equation*}
c_p(M,A)=Cost(A)+\sum_{N\in E(M)}{\cal L}\big(p_{(M,A)}(N)\big).
\end{equation*}

}

\item{\label{problem2.3}
Use the method described in the solution of Problem 1 with cost function $c_{p}$ for $\cal M,A$ to get a better assignment $p'$.}

\item{
Continue with step \ref{problem2.2} using the assignment $p'$ given in \ref{problem2.3}.}

\end{enumerate}

We run this procedure for a number of rounds.

\begin{remark}
\label{rem.endless.loop}
With no further adjustment, the algorithm can easily result in an infinite loop as the following example shows:\\
\noindent
Consider two macros and let the netlist be one single net connecting the two macros.
During the run of the algorithm if the two macros are at different position, the macro with the higher cost would move to the area where the other macro is, because this decreases the total netlength and the total cost of the macros.
The cost of this area increases until one of them moves to another place.
As before the net connecting this macro to the other causes the other macro to move as well to the same area.
Therefore the same process starts again.
This shows that increasing the cost under the overlaps alone is not enough.
\end{remark}
After each round in the algorithm if there are at least two macros at the same area, we increase the cost of that area.
At the beginning of the algorithm we allow overlaps to get a better placement, basically allowing not only better, but slightly worse placements to prevent the algorithm to get stuck early in some local minimum.
Later, we increasingly punish overlaps to prevent the loop in Remark \ref{rem.endless.loop}.
This is a kind of cooling process.
If we set the increment rate properly, the macros will have the time to distribute evenly in the placement area, with small netlength.
Later, this punishment goes to infinity, hereby enforcing a legal solution.

\bigskip
\begin{problem}
There is a set of macros $\cal M$, a netlist $\cal N$, a placement area $\cal A$ which is a discrete $n\times m$ grid and a netlength model $\cal L$ given.
Each edge length of each macro is the multiple of the edge length of the grid.
A placement $p:{\cal M}\mapsto {\cal A}$ is legal if $S(M_1,p(M_1))\cap S(M_2,p(M_2))=\emptyset$, $M_1\neq M_2$.
During the run of the algorithm we allow non-legal placements.
We only require that the final placement $p$ is legal.
Our task is to find a legal placement which minimizes
\begin{equation*}
\sum_{N\in {\cal N}}{\cal L}\big(p(N)\big).
\end{equation*}
\end{problem}

\noindent
{\bf Solution.}
Here the places are not disjoint as in Problem 2, so during the run of the algorithm the macros can overlap partially as well.
In this case we increase the cost under the intersection proportional to its size.

\bigskip
\begin{problem}
In the general setting the sizes can be real numbers.\\
\end{problem}

\noindent
{\bf Solution.}
To use the solution of Problem 3, we divide the placement area to a sufficiently fine discrete grid and use only natural numbers for approximation.
We round up the edge length of each macro to the nearest multiple of the edge length of the grid.
%
%
%
%
%

\section{Our global placer}
\label{sec.global.placer}
The algorithm receives an initial placement (e.g. random with many overlaps) and then refines it to a global placement with minimized total netlength.
\subsection{The structure of the algorithm}
Our algorithm consists of \ref{macro.increase} steps as follows:
At the beginning, we generate an initial placement, or we use the given one. Then, for a given number of rounds, we do the following.

\begin{enumerate}

\item{
\label{makro.kiv}
	We choose a macro $M$ randomly with original position $X_0$.}
	
\item{We generate $t$ possible new positions $X_1, \dots , X_t$ around its original position, with
	{\it move\_macro(M)}}

\item{

	We move the macro $M$ to the positon $X_i$ that minimizes
		\begin{equation}
		\label{macro.weight}
		{\it weight(M, X_i)} + NetLength\big(E_{X_i}(M)\big) + penalty (M, X_i),
		\end{equation}
	where $E_{X_i}(M)$ is obtained by moving the macro $M$ to $X_i$.}

\item{
\label{macro.increase}
	We increase the weights under the overlaps of $M$ with every other macros.}

\end{enumerate}
\subsection{Notations}
To discretize the problem, we consider the placement area to be a finite $n\times m$ grid. We can assume that $n=2^p,m=2^q$.
These parameters are free to choose according to the available computing resources.
Denote the size of the placement area by $A_x$ and $A_y$ (for the horizontal and vertical size).
After we set $n,m$, we divide the placement area to an $n\times m$ grid.
Our grid will consist of $nm$ squares of dimensions $x\times y={A_x\over n}\times{A_y\over m}$.
We record the cost as a stepfunction on the placement area which is constant on the cells of the grid.
In other words we define the cost on the cells of the grid and not as a function on the placement area.
Let $P_{i,j}$ be the weight under the $i$th square of the $j$th row.
Denote by ${\cal M}_{n,m}$ the set of all $n\times m$ matrices.
Define the inner product of two matrices (say $A,B \in {\cal M}_{n,m}$) as follows:
\begin{equation*}
A \star B :=\sum_{i=1}^{n}\sum_{j=1}^{m} a_{i,j}b_{i,j}.
\end{equation*}
We represent a rectangle with its top-left and bottom-right corners.
For a given rectangle $\tilde{R} = \big((x_1,y_1),\ (x_2,y_2)\big)$, $x_1<x_2$, $y_1<y_2$, we consider the slightly larger rectangle
\begin{equation*}
R=\Bigg(\bigg(\Big\lfloor \frac{x_1}{x} \Big\rfloor x, \Big\lfloor \frac{y_1}{y} \Big\rfloor y\bigg),\ 
\bigg(\Big\lceil \frac{x_2}{x} \Big\rceil x,\Big\lceil \frac{y_2}{y} \Big\rceil y\bigg)\Bigg)=
\Big((a_1 x, b_1 x),\ (a_2 x, b_2 x)\Big).
\end{equation*}
This is the smallest rectangle of the grid covering $\tilde{R}$.
From now on let every rectangle be given in the form:
\begin{equation*}
R=\big((a_1 x, b_1 y),\ (a_2 x, b_2 y)\big) = \big((a_1,b_1),\ (a_2,b_2)\big).
\end{equation*}
The characteristic function of $R = \big((a_1,b_1),(a_2,b_2)\big)$ is defined as the following $n\times m$ matrix.
\begin{equation*}
A_R =(a_{i,j})_{i,j=1}^{n,m}\, ,\ 
a_{i,j}=
	\left\{
		\begin{array}{ll}
		1 & \text{if } a_1 < i \leq a_2 \text{ and } b_1 < j \leq b_2\\
		0 & \text{otherwise}
		\end{array}
	\right.
\end{equation*}

\subsection{Data structure for the weights}
We introduce a data structure by which we can calculate (\ref{macro.weight}) in ${\cal O}\big(\log (n)\log (m)\big)$ time, and also, we can increase the cost function by a constant under any rectangle $R$ (as in Problem \ref{macro.increase}) in ${\cal O}\big(\log (n)\log (m)\big)$ time.

\begin{remark}
\label{rem.computation.time}
We have two operations on $P$.
\begin{itemize}

\item
$f(R,P)=A_R\star P$ returns the total cost under a given rectangle $R$.

\item
$g(R,P,w)$ increases each entry of $P$ by $w$ under the rectangle $R$. ($P:=P+wA_R$)

\end{itemize}
In our algorithm, we use these operations in every round, therefore, we need to compute them fast.
\end{remark}

We construct an orthogonal basis $\{B_1, B_2, \ldots ,B_k\}$ in this space ($k=nm$).
For a given rectangle $R$, in order to compute $A_R \star P$, we only need to know the products $B_i\star A_R$ for every $i=1,\ldots ,k$.
Increasing the entries under $R$ by $w$ in $P=\sum_i\alpha_iB_i$ can be done by increasing the coefficients of the expansion $\alpha_i=\alpha_i+wA_R\star B_i$.
We construct a base such that for every rectangle $R$, there are only a few basis elements which are not orthogonal to $R$, and hence the inner product can be computed in constant time.
First, consider the one dimensional array $P=(p_i)_{i=1}^{n}$ and let $n=2^p$.
Define $B_k^a$ as follows.
$$B^a_k(j)=
\left\{
		\begin{array}{ll}
		1 & (2k-2)2^{a}< j \leq (2k-1)2^a\\
		-1 & (2k-1)2^a < j \leq 2k2^a\\
		0 & \text{else}
		\end{array}\right.
$$
where $a=0,\ldots ,p,\ k=1,\ldots ,2^{p-a-1}, j=1\ldots ,n$. We also consider the basis element $B^p_1=\mathds{1}$. For example, the elements for $n=8$ are
$$
B_1^0=\left[
\begin{array}{c}
1\\
-1\\
0\\
0\\
0\\
0\\
0\\
0
\end{array}
\right]
,\ 
B_2^0=\left[
\begin{array}{c}
0\\
0\\
1\\
-1\\
0\\
0\\
0\\
0
\end{array}
\right]
,\ 
B_3^0=\left[
\begin{array}{c}
0\\
0\\
0\\
0\\
1\\
-1\\
0\\
0
\end{array}
\right]
,\ 
B_4^0=\left[
\begin{array}{c}
0\\
0\\
0\\
0\\
0\\
0\\
1\\
-1
\end{array}
\right]
,\ 
$$

$$
B_1^1=\left[
\begin{array}{c}
1\\
1\\
-1\\
-1\\
0\\
0\\
0\\
0
\end{array}
\right]
,\ 
B_2^1=\left[
\begin{array}{c}
0\\
0\\
0\\
0\\
1\\
1\\
-1\\
-1
\end{array}
\right]
,\ 
B_1^2=\left[
\begin{array}{c}
1\\
1\\
1\\
1\\
-1\\
-1\\
-1\\
-1
\end{array}
\right]
,\ 
B_1^3=\left[
\begin{array}{c}
1\\
1\\
1\\
1\\
1\\
1\\
1\\
1
\end{array}
\right]
$$

\begin{lemma}
\label{lem.number.not.ortogonal}
Let $s,t\in \mathds{N},s<t$.
For a given $R_{s,t}$, there are at most $2\log(n)$ elements of the basis $B_k^a$ for which $R_{s,t}\star B_k^a\neq 0$.
\end{lemma}

\noindent
{\bf Proof}:
It is easy to check that $B_k^a$ is an orthogonal basis in $\mathds{R}^n$.
Let $R_{s,t}\in \mathds{R}^n$ as in the Lemma:
$$
R_{s,t}(j)=
\left\{
		\begin{array}{ll}
		1 & s< j \leq t\\
		0 & \text{else}
		\end{array}\right.
$$
If $B_k^a\star R_{s,t}\neq 0$ then either (\ref{cont.corner.x}) or (\ref{cont.corner.y}) or 
both holds.
\begin{equation}
\label{cont.corner.x}
(2k-2)2^a<s\leq 2k2^a
\end{equation}
\begin{equation}
\label{cont.corner.y}
(2k-2)2^a<t\leq 2k2^a
\end{equation}
For a given $R_{s,t}$, the number of $B_k^a$ for which (\ref{cont.corner.x}),(\ref{cont.corner.y}) or both holds, is at most $2\log(n)$.
This completes the proof of the lemma.
\qed

\noindent
It is easy to compute the scalar product of $R_{s,t}$ and $B_k^a$:
\begin{equation*}
Star_x(R_{s,t},B_k^a)=-\min \big\{s-(2k-2)2^a,\ 2k2^a-s\big\},
\end{equation*}
\begin{equation*}
Star_y(R_{s,t},B_k^a)=\min \big\{t-(2k-2)2^a,\ 2k2^a-t\big\}.
\end{equation*}
Then the scalar product of $R_{s,t}$ and $B_k^a$ is:
\begin{equation}
\label{scalar}
R_{x,y}\star B_k^a=\left \{
\begin{array}{ll}
Star_x(R_{s,t},B_k^a) & \text{if only (\ref{cont.corner.x}) holds}\\
Star_y(R_{s,t},B_k^a) & \text{if only (\ref{cont.corner.y}) holds}\\
Star_x(R_{s,t},B_k^a)+Star_y(R_{s,t},B_k^a) & \text{if (\ref{cont.corner.x}) and (\ref{cont.corner.y}) holds}
\end{array}
\right.
\end{equation}
\bigskip
Now we can get a basis in $\mathds{R}^{n\times m}$ from the one dimensional case as follows.
Let
$$
B^{a,b}_{k,l}(i,j)=B^a_k(i)\cdot B^b_l(j),
$$
\noindent
where $B^a_k$ corresponds to the basis in $\mathds{R}^n$ and $B_l^b$ to the basis in $\mathds{R}^m$.
It is not hard to check that $\{B_{k,l}^{a,b}\}$ is a basis in ${\cal M}_{n,m}$.
Following the argument of Lemma \ref{lem.number.not.ortogonal}, for any rectangle $R$, there are at most ${\cal O}\big(\log(n)\log(m)\big)$ basis elements ($B_{k,l}^{a,b}$) not orthogonal to $A_R$.
Furthermore, from (\ref{scalar}), the scalar product $A_{(x_1,x_2)\times(y_1,y_2)}\star B^{a,b}_{k,l}$ can be computed in constant time.

\noindent
It is easy to see that $B_{k,l}^{a,b}$ satisfies:
\begin{itemize}

	\item{
	$\forall R$ rectangle, $\Big|\big\{B_{k,l}^{a,b}:\, A_R\star B_{k,l}^{a,b} \neq 0\big\}\Big| = {\cal O}\Big(\log (n)\log(m)\Big)$.
	Furthermore, we can find them in ${\cal O}\Big(\log (n)\log(m)\Big)$ time.}

	\item{
	$\forall R$ rectangle $A_R\star B_{k,l}^{a,b}$ can be computed in constant time.}

\end{itemize}
\subsection{Inflation}
During the run of the algorithm, the cost of crowded areas may get too high, causing that all macros will avoid that area.
Rather than waiting for the costs of all the other places to increase, we implement a cost reducer.
It will reduce the differences between the high- and low-cost areas.
We chose the method below because it can be easily implemented without further computation time.
The best rate of inflation should be adjusted.
\subsection{The {\it increase(R,value)} subroutine}

Let $\alpha=(\alpha_{k,l}^{a,b})$ be a global variable denoting the coefficients of the basis elements $B_{k,l}^{a,b}$ in the expansion of $P=\sum_{k,l,a,b}\alpha_{k,l}^{a,b}B_{k,l}^{a,b}$.
The {\it increase(R,value)} subroutine computes the scalar product of the basis elements $B_{k,l}^{a,b}$ and $A_R$, and increases the current coefficient of $B_{k,l}^{a,b}$ with this product multiplied by {\it value}.
We repeat this for all $B_{k,l}^{a,b}$:
\begin{algorithm}[H]
\caption{increase($R$,$value$)}
\begin{algorithmic}
\FOR{$\{a=0,\dots, \log(n)\}$}
\FOR{$\{b=0,\dots, \log(m)\}$}

\FOR{$\{k,l:B_{k,l}^{a,b} \star A_R\neq 0\}$}
\STATE $\alpha_{k,l}^{a,b}\ =\ \alpha_{k,l}^{a,b}\ +\  scalar(A_R,B_{k,l}^{a,b})*value$

\ENDFOR
\ENDFOR
\ENDFOR
\end{algorithmic}
\end{algorithm}
In line 3, we can find the pairs $(k,l)$ in constant time as follows.
For a given $R$ there are at most $4$ pairs $(k,l)$ such that
$scalar(A_R,B_{k,l}^{a,b})\neq 0$.
The possible pairs $(k,l)$ can be found easily from the coordinates of $R$.
Fix $a,b$ and a rectangle $R=((x_1,y_1),(x_2,y_2))$, where $x_1<x_2,\,y_1<y_2$. Let $k_i,l_j$ be such that
$$(2k_1-2)2^a<x_1\leq 2k_12^a,\ (2k_2-2)2^a<x_2\leq 2k_22^a,\textrm{ and}$$
$$(2l_1-2)2^a<y_1\leq 2l_12^a,\ (2l_2-2)2^a<y_2\leq 2l_22^a$$
holds.
The basis elements (with fixed $a,b$) possibly not orthogonal to $A_R$ are $B^{a,b}_{k_1,l_1}$, $B^{a,b}_{k_2,l_1}$, $B^{a,b}_{k_1,l_2}$, $B^{a,b}_{k_2,l_2}$.
\subsection{The {\it cost(R)} subroutine}
Here, $Round\in \mathds{N}$ is a global variable denoting the current round of the algorithm.
The cost(R) function receives a rectangle $R$ and returns the total cost of the cells inside this rectangle.
This routine uses the basis expansion for the cost matrix $P$ in order to compute the scalar product as follows:
\begin{algorithm}[H]
\caption{cost(R)}
\begin{algorithmic}
\STATE $cost\ =\ 0$;
\FOR{$\{a=0,\dots, \log(n)\}$}
\FOR{$\{b=0,\dots, \log(m)\}$}

\FOR{$\{k,l:\ B_{k,l}^{a,b} \star A_R\neq 0\}$}
\STATE $cost\ =\ cost\ +\ \alpha_{k,l}^{a,b} * scalar(A_R,B_{k,l}^{a,b})$

\ENDFOR
\ENDFOR
\ENDFOR
\STATE $cost\ =\ cost\ +\ penalty(Round,R)$;
\STATE {\bf return} $cost$;
\end{algorithmic}
\end{algorithm}
%
%
%
%
\section{Heuristics}
\label{sec.heuristics}
In this section, we discuss further parameters of the algorithm.
We make suggestions for all parameters, but these should be experimentally adjusted.

\subsection{The {\it move\_macro(M)} subroutine}
This routine returns a new possible place for $M$.
As before, let $A_x, A_y$ denote the horizontal and vertical size of the placement area $\cal A$.
The location of a macro $M$ is given by its placement coordinates $(x,y)$.
For a macro $M\in {\cal M}$, let us denote the largest and smallest possible $x$ coordinates for the macro $M$ by
$$x_{max}(M)=A_x-{size_x(M)\over 2},$$
$$x_{min}(M)={size_x(M)\over 2}.$$

We define $y_{min}(M), y_{max}(M)$ analogously.
Let $\gamma(x)=\exp\big(\log (x) \cdot U\lbrack 0,1\rbrack\big)$ where $U\lbrack 0,1\rbrack$ is a uniformly distributed random variable in $\lbrack 0,1\rbrack$.
This distribution is our heuristic choice.
The subroutine:
\begin{algorithm}[H]
\caption{move\_macro($M$)}
\begin{algorithmic}
\STATE $a\ =\ $Rand\{-1,1\}
\STATE $b\ =\ $Rand\{-1,1\}
\STATE $x_{new}=
\left\{
		\begin{array}{ll}
		x-\gamma(x+1) &\text{if }\ a=1\\
		\\
		x+\gamma(x_{max}-x) &\text{if }\ a=-1\\
		\end{array}
	\right.$
\STATE $y_{new}=
\left\{
		\begin{array}{ll}
		y-\gamma(y+1) &\text{if }\ b=1\\
		\\
		y+\gamma(y_{max}-y) &\text{if }\ b=-1\\
		\end{array}
	\right.$
\STATE {\bf Return} $x_{new},y_{new}$
\end{algorithmic}
\end{algorithm}
\subsection{The {\it penalty(Step,R)} function}
\begin{algorithm}
\caption{penalty(Step, R)}
\begin{algorithmic}
\STATE $cost\ =\ 0$
\FOR{$\{M\in {\cal M}, M\neq R\}$}
\STATE $cost\ =\ cost\ +\ c*\delta_{Step}*\text{Circ}(M\cap R)$
\ENDFOR
\STATE {\bf Return} $cost$
\end{algorithmic}
\end{algorithm}
Here, Circ$(R)$ denotes the circumference of the rectangle $R$, $c$ is a constant and $\delta_{Step}$ is a parameter.

\subsection{The {\it smooth\_edge(E)} function}
We will consider the bounding-box model only.
During the earlier stages of the optimization, when we compare different positions of a macro and we calculate the total distance of the wires, then we should consider that the positions of the neighbors are still rough. Therefore, it turns out to be useful to consider the positions of the neighboring pins with some uncertainty, namely, as distributions around their present positions. This can be expressed by using a smoothed version of the absolute value function of the difference in each coordinate. This tool was already used in the literature, it is common to approximate the bounding box model (\ref{eqn.bounding.box}) with strictly convex functions which converges to the bounding-box netlength.
One of them is the log-sum-exp function (see \cite{Chan_et_al}, \cite{Cong_Luo}, \cite{Agnihotri_Madden}):
\begin{equation*}
LSE_x(N):=\alpha\log \Big(\sum_{p\in N}\exp\big(x(p)/\alpha\big)\Big)+\alpha\log \Big(\sum_{p\in N}\exp \big(-x(p)/\alpha\big)\Big),
\end{equation*}
and $LSE(N):=LSE_x(N)+LSE_y(N)$. It is easy to see that $LSE(N)\rightarrow BB(N)$, as $\alpha\rightarrow 0$.\\
An alternative way is to approximate with $L_p$ norms (see \cite{Alpert_et_al}):
\begin{equation*}
LP_x(N):=\sum_{p,q\in N}\Big(\big(x(p)-x(q)\big)^p+\alpha\Big)^{1/p},
\end{equation*}
and $LP(N):=LP_x(N)+LP_y(N)$. $LP(N)\rightarrow BB(N)$ holds again, if ${1\over\alpha}\rightarrow \infty,\ p\rightarrow \infty$.

We used exponential functions, in a way similar to the log-sum-exp model, as follows:
\begin{equation*}
NL_x(N) = {1\over\beta} \sum_{p\in N}\log\Big(\exp\big(\beta x(p)\big) + \exp\big(-\beta x(p)\big)\Big),
\end{equation*}
and $NL(N)=NL_x(N)+NL_y(N)$. It is clear that $NL(N)\rightarrow BB(N)$ holds if $\beta \rightarrow \infty$.
We use
\begin{equation*}
\beta={MaxRounds\over MaxRounds-Round+1},
\end{equation*}
where $MaxRounds$ is the number of rounds for which we want to run the algorithm.
%
Formally the code of this subroutine is as follows:

\begin{algorithm}
\caption{smooth\_edge(E)}
\begin{algorithmic}
\STATE $C_x = {1\over\beta} \log(\exp(\beta x(E))+\exp(\beta(-x(E))))$
\STATE $C_y= {1\over\beta} \log(\exp(\beta y(E))+\exp(\beta(-y(E))))$
\STATE {\bf Return} $C_x + C_y$
\end{algorithmic}
\end{algorithm}

Notice that after many rounds, the edge length tends to the actual Bounding-box netlength.

\subsection{Possible remaining overlaps}
It is usually useful to stop the global placement before it removes all the overlaps.
Our placer is ineffective in the very final stages of the algorithm, when the actual placement is almost legal, and only a few small overlaps should be eliminated.
Therefore, we can get slightly better results if we stop the algorithm before the very final steps, and we use some other final legalization method,  even a simple naive one. In our case, these final minor modifications were performed by hand.

\section{Conclusions}

In this paper we gave a heuristic algorithm for the NP-hard macro placement problem.
The design of the algorithm is based on a primal-dual approach to a matching problem (see Section \ref{sec.basic.tools}, Problem 1).

First, we implemented a special data structure to handle the dual (cost) function efficiently during the algorithm.
This can records a multidimensional (in our case, 2-dimensional) discrete function, and performs efficiently the following two operations. It returns with the sum (integral) of the values in any rectangle, and it can increase the function with any constant in any rectangle. This data structure can also be useful for other purposes.

The second part includes the heuristics (see Section \ref{sec.heuristics}) inspired by the Hungarian Algorithm.
We suggest an algorithm that iteratively revises the primal and the dual functions.
Despite a pair of optimal primal-dual solutions do not exists, this causes problems only around the finalization of the placement.
Our this heuristics seemed to perform well for finding good rough positions for the macros.
Therefore, we used a natural continuous transition of the primal-dual method to a simple algorithm which just enforces disjointness.
There were many minor details where we found nontrivial solutions which can be used in other problems, as well.
All these together provide a flexible and robust algorithm for the VLSI placement problem, which can be easily optimized for different scenarios.

\end{document}